\documentclass{ws-procs9x6}

\begin{document}

\title{\hspace*{1.5cm}Hydrodynamic aspects \newline of  Relativistic Heavy Ion Collisions at RHIC}

\author{Peter F.~Kolb}

\address{Physik Department \\
Technische Universit\"at M\"unchen\\ 
D-85747 M\"unchen, Germany\\ 
E-mail: peter.kolb@ph.tum.de}

\maketitle

\abstracts{
The current status of the application of hydrodynamics to 
ultrarelativistic heavy ion collisions is reviewed. 
We elaborate on the arguments for strong transverse flow and
 rapid thermalization and 
discuss future applications and trends in hydrodynamics.  
}

\section{Overview}
Among the many predictions\cite{Eskola01} 
made for observables of collisions at the Relativistic Heavy Ion Collider (RHIC),
hydrodynamic calculations regarding the expected 
flow characteristics\cite{KSH00,TLS01} 
turned out to be surprisingly accurate already a few 
months after running the new facility.\cite{STARv201}
In fact, those predictions were so convincing that they were 
awarded a bottle of wine in the RHIC 
prediction competition.\cite{RHICpredictioncontest}
That RHIC creates systems which -- {\em for the dominant part
of their lifetimes} -- expand according to hydrodynamic
principles is of utmost importance:
The dynamical evolution is governed by the nuclear equation
of state at extreme energies, 
whose thermodynamic features are thus accessible through collisions
in the laboratory. 

In the following, I will review indications of the
strong transverse expansion that can be 
deduced from the observable final state 
of central and non-central Au+Au collisions. 
Thereby, I will pay particular attention to
signals of collectivity at mid-rapidity, 
i.e. in the plane perpendicular to the beam-axis 
in the center of mass of the reaction (at vanishing
pseudorapidity $\eta$). 
Although interesting phenomena are to be expected at $\eta \ne 0$, 
to which I will return later, the midrapidity region is of particular interest
for the observation of collectivity: Before the reaction, there is no 
transverse component of motion. Observing collectivity in the transverse
plane at midrapidity must thus be a result of the dynamical evolution
of the fireball {\em after} the initial impact. 
In contrast, at rapidities $\eta \ne 0$, it is difficult to 
disentangle collectivity created by the fireball region and the
longitudinal dynamics of the nuclei that was present already before
the reaction.

To address the questions of the how and when those collective 
features in the transverse  plane 
are established, I will turn to dynamical models and 
specifically introduce the formalism of hydrodynamics to study
the temporal evolution of the fireball matter in this framework. 
This analysis makes the success of the simple formulas of blast wave
parameterizations understandable and validates such 
a simplistic approach
as a tool to rapidly characterize and summarize 
the main thermodynamic information of the final state.

I will then highlight the most important comparisons of experimental
data and hydrodynamic calculations and their interpretation, 
before I give a detailed account 
on recent and future developments that can be expected in the field. 
Almost every single observable of heavy ion physics is 
influenced by the bulk matter, as the bulk provides the background of the 
evolution through which even the rare probes have to propagate.
Now, with hydrodynamics, we have a tool to calculate the dynamics of this
evolution.

\section{Indications of strong transverse flow}

If the microscopic constituents of a flowing medium share a common
average flow velocity $v_\perp$, 
heavier particles in this medium gain larger momenta than
the lighter constituents. In the case of the late stages of heavy ion collisions, 
the flowing medium is the hadronic soup consisting of hadrons of widely varying
masses. From experimental evidence we find that the hadronic abundancies
reflect a composition of a hadronic gas at a temperature
$T_{\rm chem} \sim 175\,{\rm MeV}$.\cite{BMMRS01} 
Refinements of the simple {\rm blast wave model}\cite{SR79,SSH93} assume
the knowledge of a certain flow profile at the break up stage of the reaction
as well as the shape of this surface of decoupling from which particles 
supposedly do not interact any further. 
In this way one achieves a handy formalism,\cite{BF01} that can easily be used
to fit to the vast amount of data currently available from RHIC experiments, in 
particular to single particles spectra and eventually their anisotropies.\cite{RL03} 
Although the temperature parameter of these fits differs widely from 175 MeV \cite{BF01}
to 100 MeV,\cite{RL03} the common finding of these model calculation is that 
a vast amount of transverse flow
(with an average velocity greater than $c/2$) is achieved on a rather short
time scale of about 10 fm$/c$. While the stunningly good description hints that there
possibly is some physical truth underlying these parameterizations, 
these simple descriptions of the final 
state do not offer any insight {\em how}  and {\em when} such strong transverse
flow develops during the course of the evolution. For this, we have to turn to a fully
dynamical description of the transverse expansion stage of the reaction.

\section{Hydrodynamic formalism and phenomenology}

There are two main philosophies to model the dynamical evolution of a system.
If the system is small and the scattering processes can be treated individually,
one can adapt a microscopic viewpoint, treating all the collisions of the constituents
individually.\cite{MG00} 
In a very dense, strongly interacting medium, however, this approach becomes quickly
impractical, particularly if one also were to consider low-momentum transfer processes
consistently. 
Alternatively, 
in this limit of high density and strong rescattering, one can give up the 
particles' individual personality, and characterize the system in terms of density fields
and continuity equations. Ideally, if scattering is strong enough to allow for local
thermodynamic equilibrium, the equations of energy momentum conservation can 
be formulated by the thermodynamic equivalent, the conservation of the
thermodynamic energy-momentum tensor,
$\partial_\mu \left[ (e+p) u^\mu u^\nu - p \, g^{\mu \nu} \right] = 0$\,. The 
energy density $e$ and the pressure $p$  are related by the thermodynamic
equation of state of nuclear matter. $u^\mu$ is the collective four velocity, whose time evolution
we can study within this formalism. Once supplied with appropriate initial conditions, 
which are often taken from geometric considerations\cite{KHHET01} or more 
fundamental models such as  the saturation model\cite{KHHET01} or the color 
glass condensate,\cite{HN04}
the system's evolution is fully determined. 
At this point, the computer takes over to solve for the time evolution of the 
thermodynamic fields.

\begin{figure}[htbp] 
\centerline{\epsfxsize=5.6cm\epsfbox{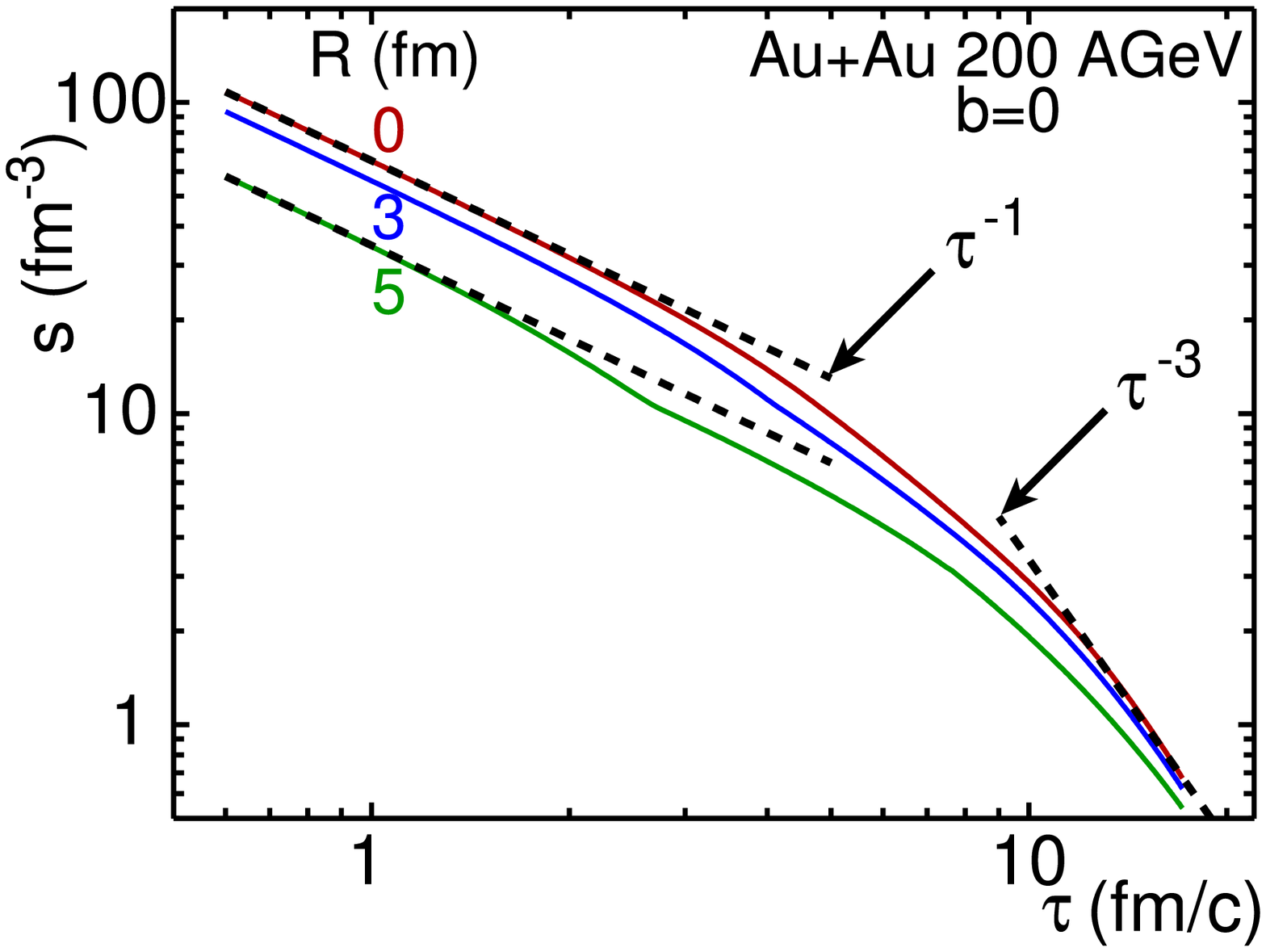}
            \epsfxsize=5.6cm\epsfbox{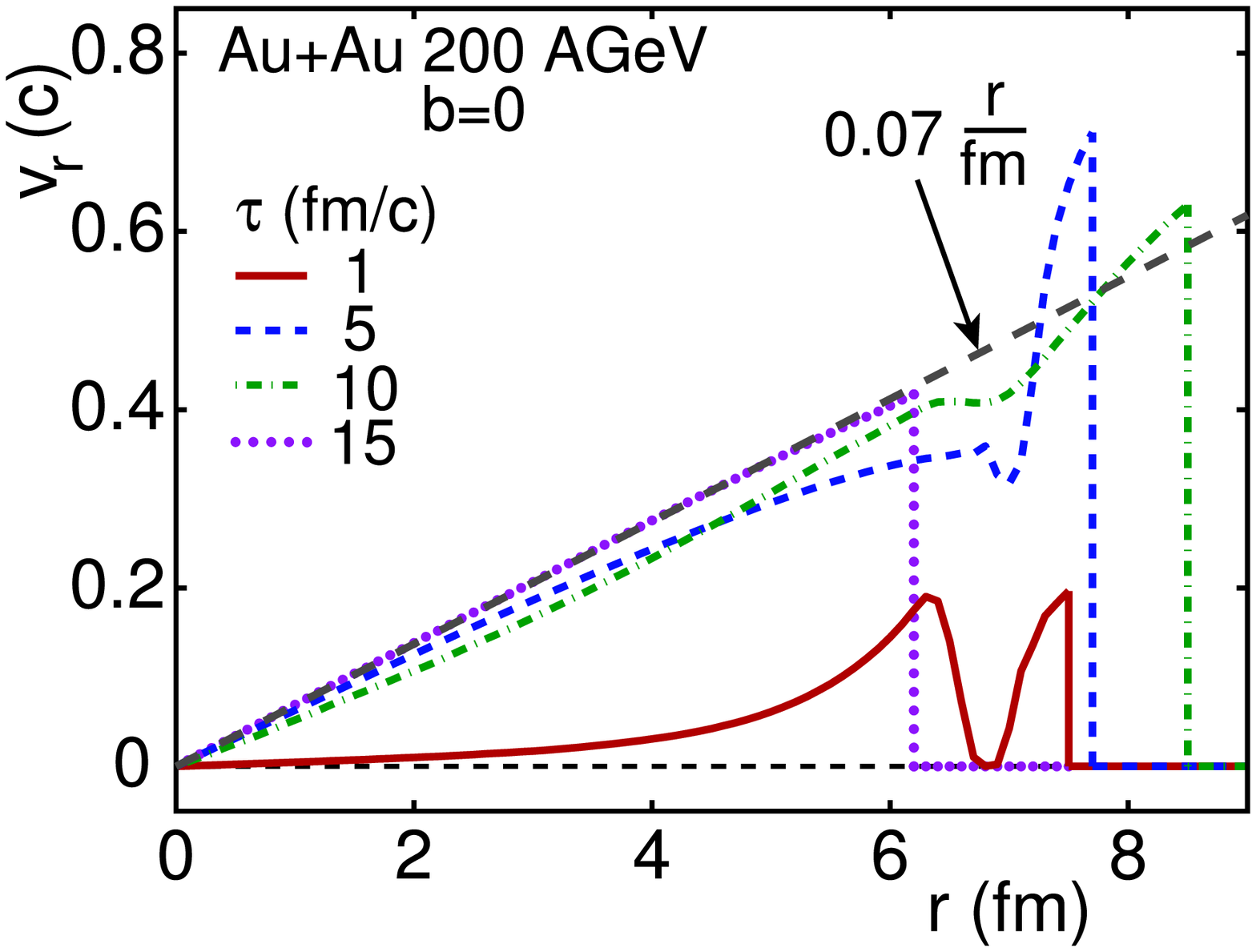}}
\caption{Left panel: Time evolution of the entropy density at three 
         different points in the fireball (0, 3, and 5~fm from the center).
         Dashed lines indicate the expectations for pure one-dimensional 
         and three-dimensional dilution, respectively.\protect\cite{Kolb03}
         Right panel: The radial velocity profile at different times during the
         evolution.\protect\cite{Kolb03} Shown is also the linear profile $v(r) = 0.07~r/{\rm fm}$.
\label{Fig1} 
} 
\end{figure}

Fig. \ref{Fig1} shows the temporal evolution of the entropy density
on a double logarithmic scale for three different positions in the fireball
of a central collision, 0, 3 and 5 fm away from the fireball center.\cite{Kolb03} 
The tangents drawn to the curve stress the occurring transition 
from the initial one-dimensional, longitudinal expansion to a fully 3-dimensional
expansion at late stages. The right plot of Fig.~\ref{Fig1} shows the radial flow
profile at different times throughout the evolution. Striking here is the fast transition
to a linear profile, which persists throughout time. 
Still, as matter is continually moved to larger radii, the mean transverse velocity
of the medium increases and transverse flow thus continues to increase. 
The main part of the fireball matter is know to freeze-out over a rather short time
frame as the rescattering rate drops sharply as a function of temperature, and therefore
even more so as function of time. 
This fact, together with the rapid development of a flat flow profile is at the
heart of the applicability of the simple and the improved blast wave parameterizations.

%
%

In contrast to the overall radial flow which increases throughout the lifetime of the system,
anisotropies in the transverse flow profile are generated during the earliest instances of
the reaction.
Such anisotropies arise from the eccentricity in coordinate space of non-central collisions. 
Larger pressure gradients in the short direction (the direction of the impact parameter)
lead to a larger transport of matter 
in this direction, thereby reducing the eccentricity and undermining the source of its own
origin.\cite{Sorge97}
In the course of this process more particles are transported into the direction of the impact
parameter and the mean transverse momentum of the particles is greater, 
both of which manifests
itself experimentally in anisotropies $v_n$ of the transverse momentum spectra
$
\frac{dN}{p_T dp_T dy d\varphi}
=
\frac{dN}{2 \pi p_T dp_T dy} \left (1 + \sum_n v_n \cos (n \varphi)\right)\,.
$
It has been shown in many studies, both microscopically\cite{ZGK99} as well as
macroscopically\cite{KSH00,Kolb03}
that those momentum anisotropies are generated during the first 5 fm/$c$
of the reaction, where most of the matter is still at temperatures exceeding
the critical temperature of nuclear matter. Anisotropies in the
transverse momentum spectra of hadrons thus originate from the partonic
stage of the fireball and signals their collective behavior.\cite{KSH00}

\section{Application to experimental observations}

As mentioned before, the transverse momentum distribution at moderate $p_T$ 
are the natural choice of observables to study collective dynamical effects in the medium.
We use the transverse momentum spectra of identified particles to 
determine the parameters of the calculation for RHIC at 200 GeV center of mass
energy per nucleon. Those spectra are well reproduced applying an equilibration time of 
$\tau_{\rm equ} = 0.6 \,{\rm fm}/c$ and a central fireball temperature of 360 MeV.\cite{KR03}
Although the particle ratios reflect chemical equilibrium at a temperature of about 
175 MeV,\cite{BMMRS01} the slopes in particular of heavier baryons require 
further transverse accelertion down to temperatures of about 
100 MeV.\cite{KH03} 
To keep the 
particle ratios fixed to their value at 175 MeV, 
chemical potentials seem to be dynamically generated throughout the
evolution.\cite{Rapp02,Teaney02,HT02}

%
\begin{figure}[htbp] 
\centerline{\epsfxsize=5.6cm\epsfbox{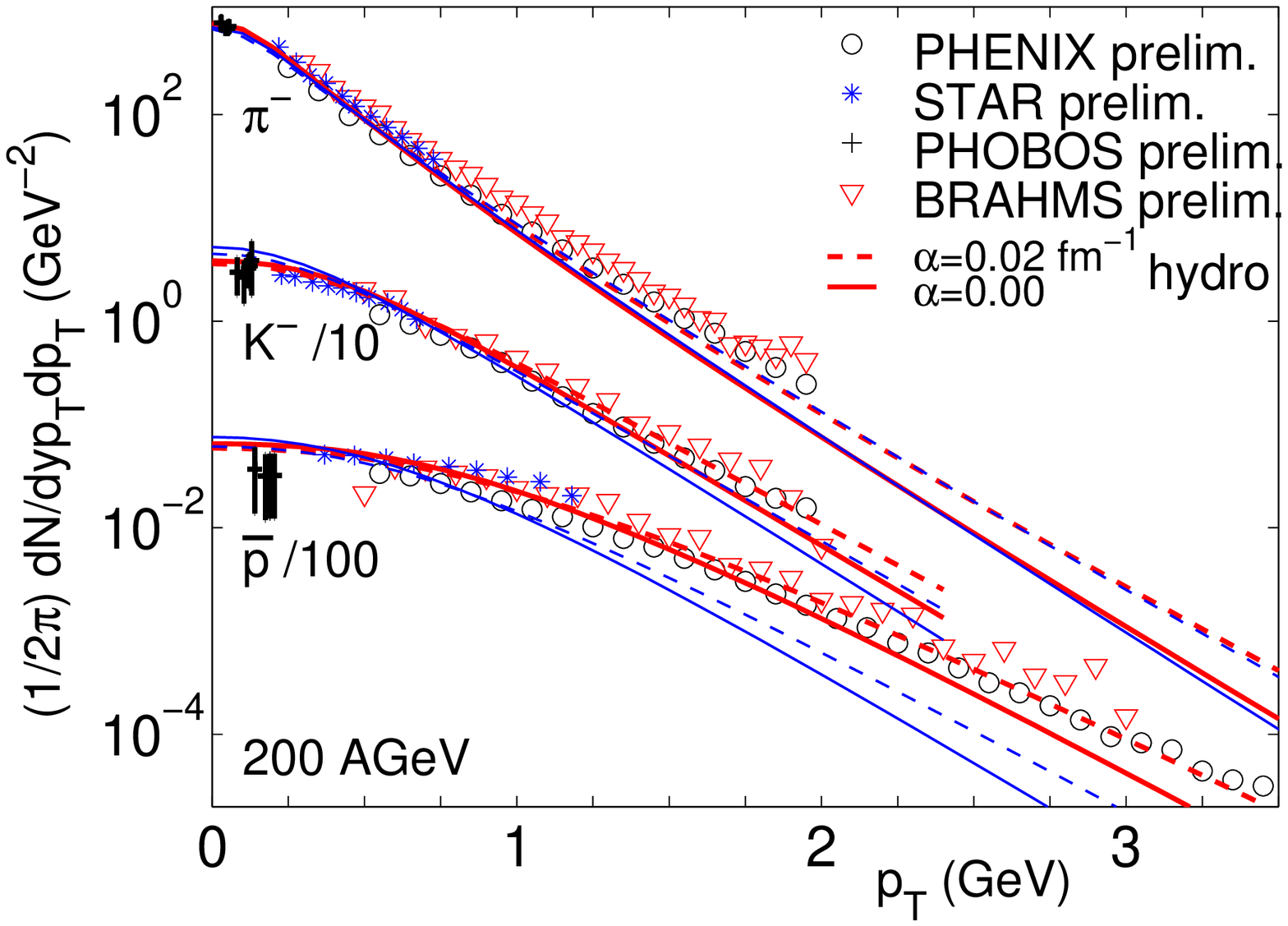}
            \epsfxsize=5.6cm\epsfbox{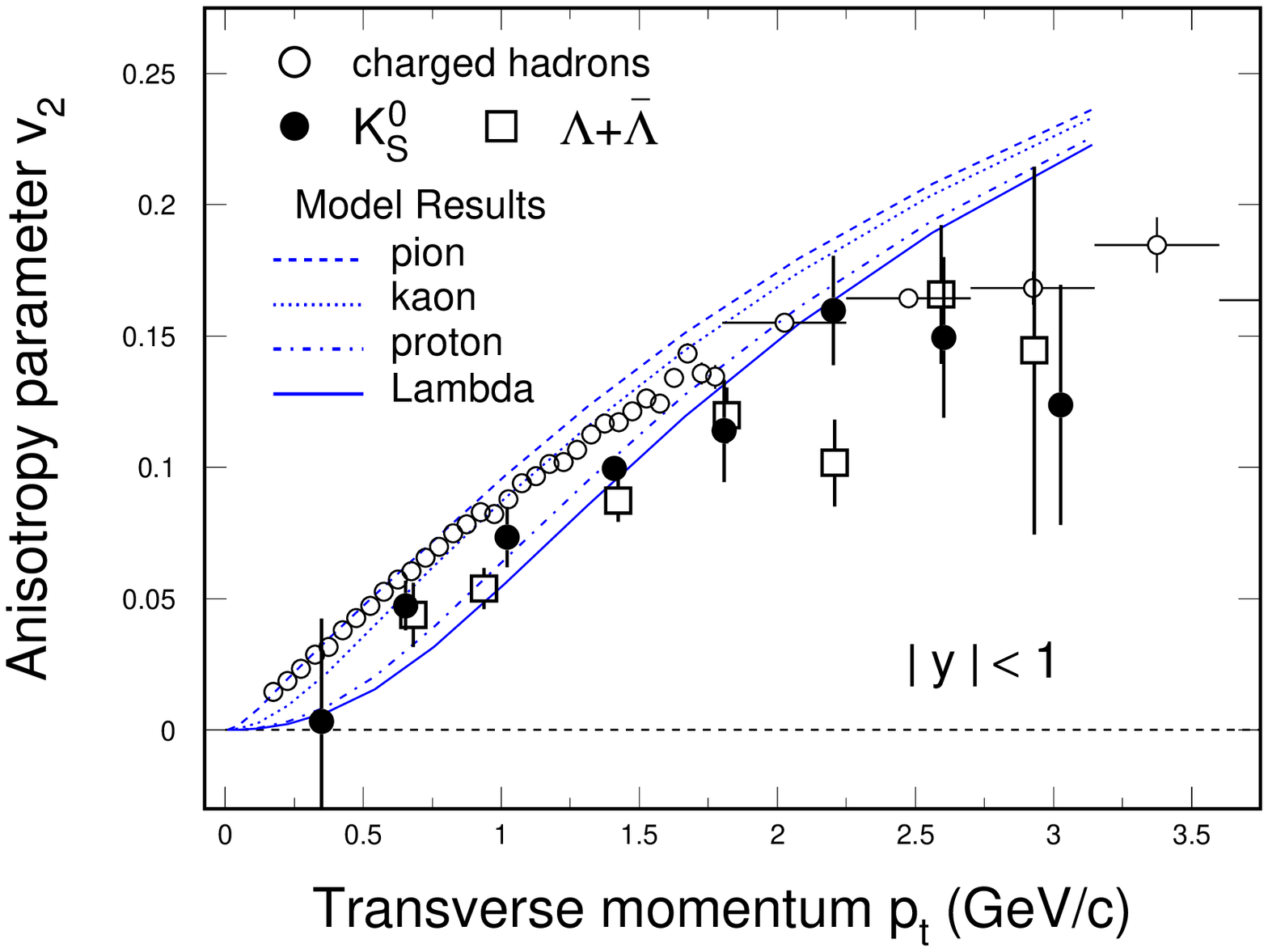}}
\caption{Left panel: Transverse momentum spectra
        of Pions, Kaons and Antiprotons, from all four RHIC experiments
        combined\protect\cite{RHICspectra}
        in comparison to results from hydrodynamics at different freeze-out temperatures
        (and different initilizations, see\protect\cite{KR03}).         
        Right panel: Momentum anisotropies of Pions, Kaons and Lambdas. Results
        from the STAR experiment\protect\cite{STAR02v2KLambda}
         compared to results from hydrodynamics.
        \label{Fig2}
        }
\end{figure} 
%
Although the good reproduction of the spectra of central collisions is reassuring, 
it has been achieved to a great deal by adjusting the parameters of the calculation. 
Important {\em predictions} come into play in non-central collisions. 
Changing centrality, which introduces a breaking of the  azimuthal symmetry inherent in 
central collisions, renders essentially  all observables sensitive on the azimuthal angle of observation.
This literally opens up an entirely new dimension of observable space.
However, for fully dynamical simulations no additional parameters are introduced
when stepping out to explore observables in this new dimension.
The modified initialization for non-central collisions is 
stricktly determined by the geometry of the underlying initialization scenario. 
The only appearing {\em parameter} -- the impact parameter $b$ -- is estimated
experimentally by giving the range of centrality for certain collisions.

We already mentioned the great significance of momentum anisotropies as they 
provide a signal from the first few fm/$c$ of the collision. 
Microscopic studies\cite{MG00}
in comparison with experimental data
show that the fireball constituents must undergo an incredible amount of rescattering 
among themselves. Hydrodynamic calculations, which exploit the limit of negligible
mean free pathlengths, provide an impressive account of a large collection of
momentum anisotropies.\cite{KH03}
 They give a good overall description of momentum anisotropies 
up to transverse momenta of 2 GeV (Pions) and even 4 GeV for Baryons. 
The centrality dependence of the hydrodynamic elliptic anisotropy compares well 
with experimental data as long as the impact parameter does not get too large. 
Finally, the mass characteristics,
a flatter onset of momentum anisotropies at small
transverse momenta,\cite{HKHRV01} is confirmed by experiment in stunningly good agreement. 
A recent compilation of momentum anisotropies of heavy particles is reprinted in
Fig. \ref{Fig2} (right panel).\cite{STAR02v2KLambda}
Still more recently the momentum anisotropies of Cascades and Omegas have become
available.\cite{Castillo04} 
Again, those anisotropies have been stunningly large, as predicted 3 years ago by 
hydrodynamic calculations.\cite{HKHRV01}
As the multistrange resonances are believe to not interact strongly in a hadron gas, this
is the clearest signal that strangeness anisotropy is generated in the partonic stage of the 
reaction. It appears that all quark flavors share the same flow anisotropy on the partonic 
level. This fact could be most prominently (dis-)proved by investigating momentum
anisotropies of higher order and their ratios.\cite{v4}
That the momentum anisotropies are as large as observed is a strong indication that 
the thermalization of the medium is achieved very rapidly.\cite{thermalization}
The fact that hydrodynamics overestimates anisotropies at large transverse momenta
and in  peripheral collisions is well understood, considering that high $p_T$ particles escape
the fireball too rapidly to follow the collective motion of the bulk, and that small systems do
not provide enough rescattering in the limited volume. Here, it is also important to remark that while 
hydrodynamics does not seem to be fully applicable at the lower beam energies of the
SPS in peripheral collisions, there might still be a fair chance for its quantitative application
in {\em central} collisions. This could be in particular true for the early, plasma part of the reaction, 
whereas the later hadronic stage might drop out of local thermal equilibrium. Only detailed
quantitative comparisons of hydrodynamic calculations and the most recent 
results\cite{NA49}
 from
SPS experiments can answer this question and address important topics close to thermal
equilibrium and just above the transition temperature.

\section{Current trends and future requirements}

Most of the hydrodynamic calculations that are applied to experimental data at present 
include some great simplifications in order to keep the numerical efforts to a reasonable 
degree. 
One of these is the strong bias to investigations of observables at midrapidity. Although
the collectivity observed at midrapidity is clearly produced during the evolution of the
produced fireball and therefore contains clean probes of the dynamics, a broader view of 
the full collision volume is highly desirable, in particular in the light of recent data 
extending the analysis of momentum anisotropies to large rapidities.
Fully three-dimensional calculations have been successfully performed 
in recent years.\cite{MMNH02}
However not too much is gained by applying the {\em ideal} hydrodynamic 
calculations at large forward rapidities as the system is thought reach thermal 
equilibrium only at later times\cite{KH04} or not at all\cite{Hirano01}
in regions far away from the center of mass of the reaction.

Nevertheless, there is a lot to be gained from investigations of
hydrodynamic calculations at non-zero rapidity. Due to the (net)-baryon content 
of the initial nuclei, one can expect that the (net) baryon-number density 
increases with increasing rapidity. As one leaves the 'clean' midrapidity
region with its antibaryon to baryon ratio of about 0.75, one gets closer to the initial
nuclei region where the antibaryon content effectively vanishes. Thus, at larger rapidity,
a hydrodynamical system evolves according to the equation of state at larger 
baryon chemical potential than at mid-rapidity. This offers the exciting possibility 
of tracing signals of the expected tricritical point at finite rapidity.
The effect of the tricritical point on the hydrodynamic evolution of RHIC collisions is
currently 
explored by Nonaka.\cite{NonakaBNL03} It was found that such a point acts as an attractor 
for the thermodynamic paths of a constant fraction of entropy and baryon-density.
Clearly, due to the large fluctuations that appear around the critical point, the 
ideal hydrodynamic treatment will become invalid, but it will be interesting to explore
how the system drops out of equilibrium, and which quantitative experimental 
measurements will be predicted. 

\begin{figure}[htbp] 
\centerline{\epsfxsize=5.0cm\epsfbox{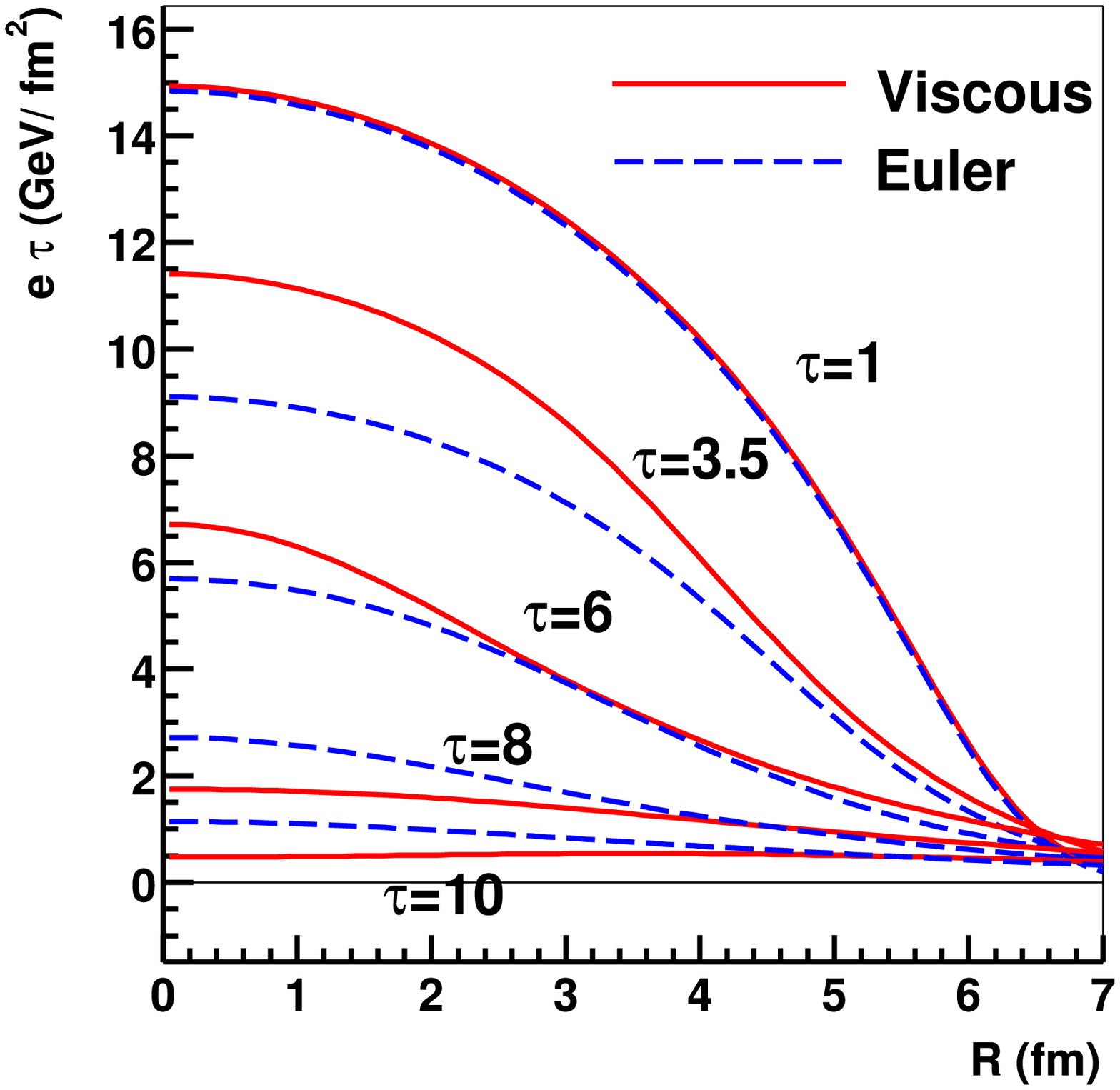}
            \epsfxsize=5.6cm\epsfbox{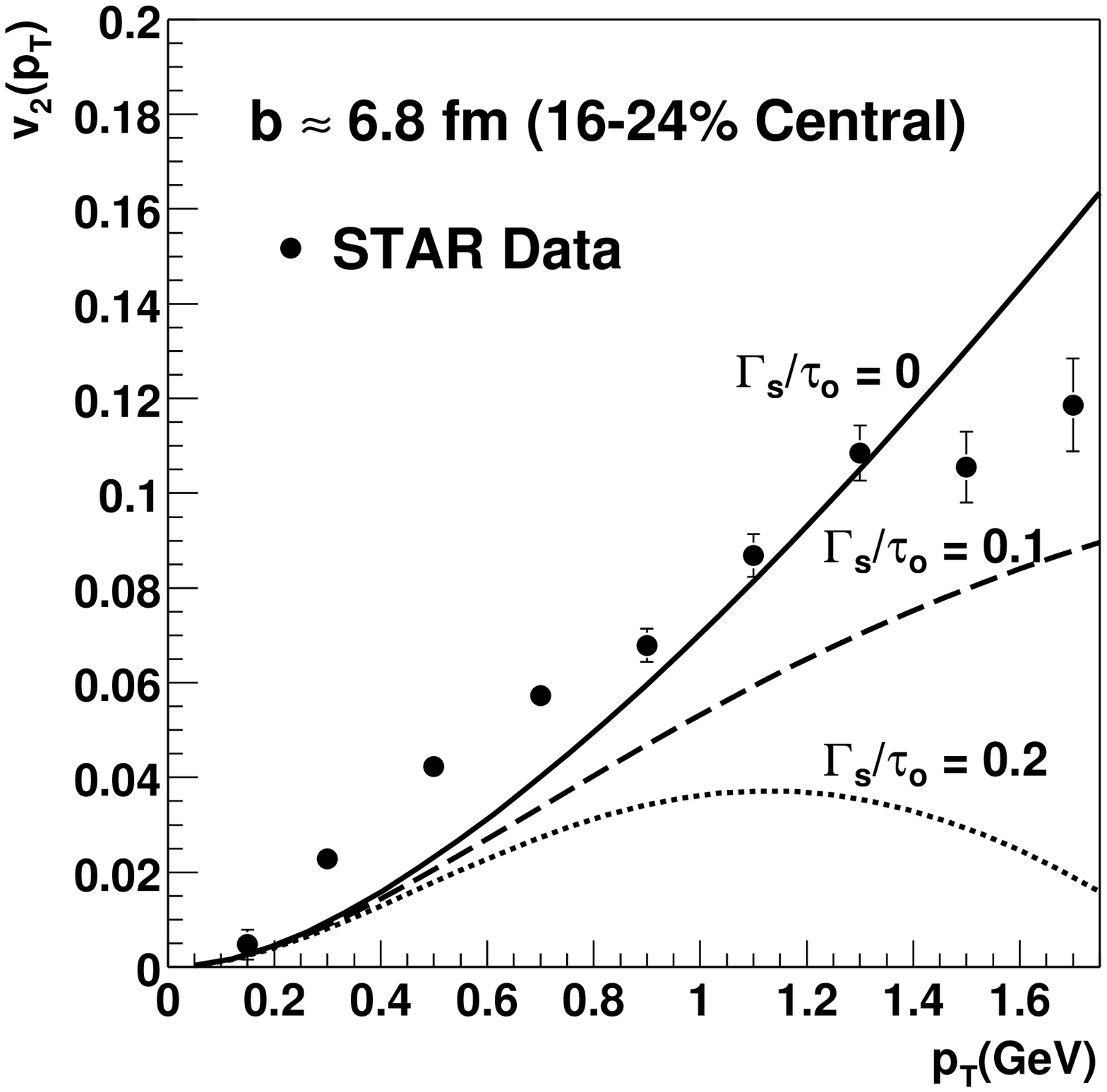}}
\caption{Left panel: Radial change of the energy density at various times
        in the ideal calculation and in a viscous calculation.\protect\cite{Teaney04}
         Right panel: Changes on the final state momentum anisotropy due to 
         deviations from the ideal of momentum space distributions.\protect\cite{Teaney03}
\label{Fig3} 
}
\end{figure}

The less efficient energy density production and deposition at large rapidity 
eventually leads to a breakdown of the assumption of full thermalization and
an ideal expansion. Viscous effects will become more and more important in
smaller and more dilute systems. It is of fundamental interest to get estimates of 
the viscosity of the QCD plasma state. Viscous hydrodynamics can deliver 
valuable information in this direction.
The realization of fully dynamical hydrodynamical calculations is however 
difficult. The viscous terms have to be solved dynamically, increasing the
number of differential equations by up to 5 (depending on the symmetries 
one exploits explicitly). Although difficult to solve,\cite{Muronga04}
first results on the dynamical evolution of central collisions have recently been
presented.\cite{Teaney04}
The message of these dynamical studies is that viscous matter sticks together for 
a longer period, just to blow up quicker in the late stages. 
The left panel of Fig. \ref{Fig3} shows plots of the energy density
as a function of the transverse coordinate for a viscous calculation compared to
the ideal Euler calculation.\cite{Teaney04}
Although the densities change slower at first, velocities build up faster due
to a restructuring of the pressure components, which is also observed in the case of
only transverse thermalization and longitudinal free streaming.\cite{HW02}
Finally, in the late stage of the reaction, the left over matter dissipates more rapidly,
due to the larger velocities that were achieved.
The net result is that the system stays longer in the hot phase of its reaction and 
disperses faster through the late stages. The late stages, for which viscosity become
increasingly important, thus have a shorter lifetime and freeze-out occurs more rapidly.
Comforting is the observation that viscosity does not lead to self-enhancing features!

%
%

In this context it is also interesting to cast doubt on the assumption of having particles
distributed in momentum space according to the simple ideal distribution laws of Fermions
and Bosons. It has been shown\cite{Teaney03} that a large deviation from these
distributions smears out the directed collective motion observed in the momentum anisotropies
and is thus not supported by the data (see right panel of Fig. \ref{Fig3}.)

\section{Summary}

We have learned {\em a lot} about the collective phenomena exhibited in RHIC data
of the first few years. The data clearly show many characteristics of rapid
thermalization and an extended stage of hydrodynamic expansion of the fireball into the
surrounding vacuum. On a linear timescale, the hydrodynamic stage is  the most 
significant stage of the reaction, thereby influencing many other relevant observables.
Hydrodynamics is a clean (meaning well defined) tool, with only a small number of parameters and
easily reproducible by anybody having access to a computer. 
The two most important facts learned from hydrodynamics and the experimental data 
so far is the need for rapid thermalization (in a time of less than 1 fm/c) where matter
obtains a large transverse push due to a hard equation of state ($p \sim e/3$).
Many quantitative questions are still unanswered: 
What is the frame of error around the equilibration time, what different kinds of 
equations of state and phase transitions
would still account for the data, how does one feed back on the other?
What is
the role of viscosity in the expansion? Why does viscosity of high temperature QCD matter seem to
be so low as to allow for an ideal hydrodynamic description of the data?

\section*{Acknowledgements}

I would like to thank the organizers for inviting me to present this overview. 
I deeply acknowledge the contributions of my collaborators and friends without
whom I would not have received this honor.

\end{document}